\documentclass[12pt]{article}
\usepackage{color,amsmath,amssymb,exscale,epsfig}
\input epsf
\newcommand{\m}[1]{\marginpar{{\tiny *}} }

\def\bea{\begin{eqnarray}}
\def\eea{\end{eqnarray}}

\begin{document}
\oddsidemargin -0.3cm
\vspace*{-2.3cm}
\begin{flushright}
\normalsize{
SLAC-PUB-15035
  }
\end{flushright}

\hfill
\vspace{20pt}
\begin{center}
{\Large \bf  Enhancing the sensitivity to New Physics in the 

$t\bar t$ invariant mass distribution}
\end{center}

\vspace{15pt}
\begin{center}
\large{Ezequiel \'Alvarez$^{a,b,}$\footnote{sequi@unsl.edu.ar}}

\vspace{20pt}
\textit{$^a$ CONICET, INFAP and Departamento de F\'{\i}sica, FCFMN,\\ Universidad Nacional de San Luis \\
Av. Ej\'ercito de los Andes 950, 5700, San Luis, Argentina}

\textit{$^b$ SLAC National Accelerator Laboratory, 2575 Sand Hill Road, Menlo Park, CA 94025, USA}

\end{center}

\begin{abstract}
We propose selection cuts on the LHC $t\bar t$ production sample which should enhance the sensitivity to New Physics signals in the study of the $t\bar t$ invariant mass distribution.  We show that selecting events in which the $t\bar t$ object has little transverse and large longitudinal momentum enlarges the quark-fusion fraction of the sample and therefore increases  its sensitivity to New Physics which couples to quarks and not to gluons.  
%We find that systematic error bars play a fundamental role in avoiding the missleading behaviour of statistic error bars going to zero as the luminosity increases.  
We find that systematic error bars play a fundamental role and assume a simple model for them.  We check how a non-visible new particle would become visible after the selection cuts enhance its resonance bump. A final realistic analysis should be done by the experimental groups with a correct evaluation of the systematic error bars.
\end{abstract}

\newpage

\setcounter{footnote}{0}

The Standard Model (SM) of particle physics is a highly predictive theory which seems to be completely verified with the recent discovery \cite{d} of a particle which could be the long-sought Higgs boson.  In the forthcoming years the LHC will be devoted to check if this particle is the SM Higgs boson, and to search for New Physics (NP) which --from the theoretical point of view-- should be expected to be at the TeV scale to solve the hierarchy problem of a light Higgs.

The top sector, mainly because of the large top quark mass, but also because of its relatively little exploration, it is a preferred sector where to expect signals of NP \cite{hills}.  Recent experimental results from Tevatron \cite{cdfviejo,tevatron} could be indicating NP effects in the $p\bar p \to t\bar t$ forward-backward asymmetry, where the main production mechanism is through quark-fusion, $q\bar q \to t\bar t$.  Although many theoretical proposals \cite{afbth} have been issued in this direction, there has not been any kind of confirmation of such a signal in the related observables at the LHC \cite{aclhc,cms6}.  

The LHC is a machine which delivers huge amounts of data, but for different reasons, it is very hard to see a NP signal over the background.  Although the main reason for this difficulty is usually its hadronic character which makes the QCD background a central problem, in this letter we address a  different issue which concerns the production mechanisms for $t\bar t$ pairs.  At the LHC, $t \bar t$ production comes from $pp\to t\bar t$ and therefore is mainly driven by gluon-fusion production ($gg\to t\bar t$ accounts for $\sim 78\%$ at the 8TeV LHC).  If there would be a NP contribution which couples exclusively to quarks, then it will be diluted because of the little proportion of quark-fusion events.  If the NP contribution is already small compared to SM quark fusion $t\bar t$ production, then its effects will be highly suppressed in a raw $t\bar t$ sample.  The goal of this note is to enhance the sensitivity in the search of NP by selecting the sample in such a way that the quark-fusion fraction of the sample is increased.

\begin{figure}[!htb]
\begin{center}
\includegraphics[width=.7\textwidth]{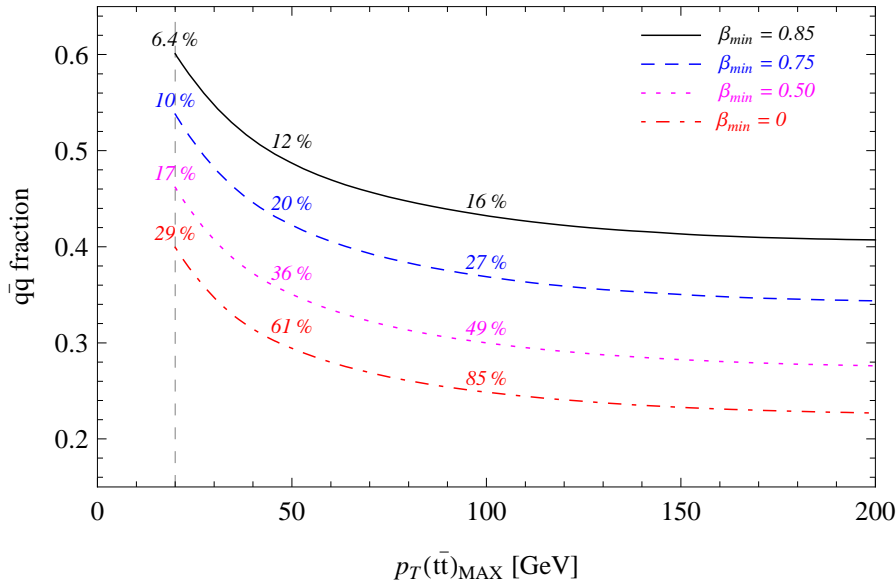}
\end{center}
\caption{$q\bar q$ fraction in $t\bar t$ production at the 8TeV LHC as cuts in $p_T(t\bar t)<p_T(t\bar t)_{MAX}$ and $\beta>\beta_{min}$ are applied.  The percentage numbers in the plot indicate the part of the original sample that passes the cuts at that point. The vertical dashed line at $p_T(t\bar t)_{MAX}=20$ GeV indicates the maximum experimental resolution \cite{cms6}.
%In all cases we also take $\beta<0.95$.
}

\label{qqbarfraction}
\end{figure}

There are mainly three simple kinematic features which can be used to increase the ratio of quark- to gluon-fusion events in $pp \to t\bar t$.  The first one comes from the proton parton density functions (PDF's): since valence quarks are more likely to have more momentum than the gluons and the sea quarks, then the events in which the $t\bar t$ pair is boosted along the beam axis have an incremental probability of coming from quark-fusion production \cite{chinos,as}.  This may be quantified through the variable \cite{as}
\bea
\beta = \frac{|p_t^z + p_{\bar t}^z|}{E_t+E_{\bar t}},
\label{beta}
\eea
where $p_{t/\bar t}^z$ is the momentum of the $t/\bar t$ quark in the $z$ direction and $E_{t/\bar t}$ its energy. $\beta$ ranges from 0 to 1, and as $\beta \to 1$ the boosting of the $t\bar t$ pair along the axis is increased and hence also the probability of coming from quark-fusion. The second kinematic feature which helps to recognize quark-fusion events is the transverse momentum of the $t\bar t$ pair, $p_T(t\bar t)$: since hard gluons in the initial state are more likely to emit initial state radiation (ISR) than quarks, then events with small $p_T(t\bar t)$ have an incremental probability to have been produced through quark-fusion.  The reason for this ISR differentiation comes from the QCD Lagrangian, where the color factor in gluon radiation from a gluon line is $9/4$ times larger than from a quark line.  This property has been previously used in top physics at the LHC \cite{german, yo} and also in Higgs physics \cite{deflo}.  The third variable which can be used to enhance the quark-fusion content of a sample is the $t\bar t$ invariant mass,  $m_{t\bar t}$, however since the purpose of this work is to study the $m_{t\bar t}$-spectrum we will not use this variable. There are other variables \cite{yang,ucla} concerning top polarization and spin correlation which are harder to implement from the experimental point of view, but could eventually also help to differentiate quark- from gluon-fusion.  

To illustrate how the first two features mentioned above are implemented in a $t\bar t$ sample,  we have plotted in Fig.~\ref{qqbarfraction} how the quark-fusion fraction is modified as different combined cuts in $\beta$ and $p_T(t\bar t)$ are applied to an original LHC $t \bar t$ raw sample.  We also point in the figure the strength of the cut with respect to the original sample.  The plot gives a good description of the expected $t\bar t$ sample as a function of the relevant variables of the problem.  The simulation in the figure is at parton-level showered by \texttt{Pythia} \cite{pythia} to include the ISR effects.  
%In this plot, and along all this work, we have always performed a $\beta<0.95$ cut, since events with $\beta$ above this cut at parton level will be hard to reconstruct at the experimental level.	

Given the previous paragraph discussion we now investigate how a selection that increases the quark-fusion fraction could enhance the sensitivity to NP in the study of the $m_{t\bar t}$-spectrum.  This kind of selection will enhance the NP contribution if the new particle couples to quarks and not to gluons.  Many of the NP proposals to solve the Tevatron $A_{FB}$ puzzle will give a contribution with these features at the LHC. In particular,  if the NP is in an $s$-channel then a resonant peak will be enhanced over the background. 

In order to study NP effects in the $m_{t\bar t}$-spectrum, we have simulated $t\bar t$ production for the SM and for a benchmark resonant NP model.  We have used \texttt{MadGraph5} \cite{mgme} to simulate the 2012 expected $t\bar t$ production for $20$ fb$^{-1}$ at the 8TeV LHC.  We have showered the parton level outcome with \texttt{Pythia} \cite{pythia} to include the initial state radiation. To avoid double counting, we have matched the matrix element with up to one extra jet to the parton shower through the MLM scheme \cite{mlm} implemented in \texttt{MadGraph5}.  We have used a K-factor of 1.55 which comes out from comparing the same simulation at 7 TeV to $t\bar t$ production at NNLL order \cite{ttbarth} at this energy.  We have assumed an overall 6\% selection efficiency \cite{cms6} to estimate the number of available $t\bar t$ semileptonic events by the end of the 2012 run.  

As a benchmark model we have taken a gluon prime resonance of mass $m_{G'}=700$ GeV, couplings to the right top $f_{t_R}=4 g_s$ ($g_s$ the strong coupling) and to the other right quarks $f_{q_R}=-0.06 g_s$, whereas all other couplings are set to zero.  These couplings yield a width $\Gamma_{G'}=90$ GeV.  This benchmark model comes from the allowed points in Ref.~\cite{julio} and besides of successfully explaining the CDF forward-backward asymmetry \cite{cdfviejo} also passes the constraints imposed by the LHC charge-asymmetries \cite{aclhc,cms6} and, due to its small coupling to light quarks, the dijet NP searches \cite{dijet}.  The production cross section for this resonance is $1.0$ and $1.3$ pb at the 7 and 8 TeV LHC, respectively.  These values should be compared to the $50$ pb production cross section at 7 TeV of the resonance which is ruled out in Ref.~\cite{cms-11-009} through a $m_{t\bar t}$-spectrum bump search.

We have computed and compared the $m_{t\bar t}$-spectrum for the SM and the NP benchmark model.  We have divided both spectrum with $25$ GeV bins and compared them through a $\chi^2$ test in a region that includes the $m_{G'}$ resonance.  We have first used only statistic error bars and then included a simple model for systematic error bars to avoid the misleading behaviour of tiny error bars due to the large sample.

If only statistic error bars are taken into account, we have found that the increase in the statistic error bars due to the selection cut spoils the visibility of the NP bump.  This happens even though the NP bump is enhanced by the selection cut.  However, this is not realistic, since given the large number of $t\bar t$ pairs expected in the LHC one is hardly ever in a pure statistic regime and systematic error bars are required as well in the analysis.  %Moreover, in the study we present here they are found to be crucial.

Since systematic error bars are found to be decisive in this analysis, a final state realistic analysis could only be performed through a correct computation of theoretical and detector systematic uncertainties by the corresponding experimental group.  We confine in this letter to use at parton level a simple model for systematic error bars based on experimental data, and show that the proposed selection cut will in principle enhance the sensitivity in the search for NP.  Only the real experimental analysis could quantify the expected enhancement in the sensitivity of the $m_{t\bar t}$-spectrum in the search of NP.

To model the systematic uncertainties we use the rate-changing systematic error bars pointed out in table $2$ of Ref.~\cite{cms-11-009}, where the 2011 $m_{t\bar t}$-spectrum is investigated.
% which include $t\bar t$ production and its backgrounds.  
We find, for the case of one $b$-tagged jet, that these systematic error bars account for a $\sim 20\%$ systematic rate-changing error in the data.
%, which is the reference background to which the signal should be compared.  
There is no simple way to simulate the shape-changing error bars indicated in Ref.~\cite{cms-11-009}.  Given that we expect the rate- and shape-changing systematic errors to be reduced, we model in this letter a $20\%$ total systematic error bars in the 2012 expected $t\bar t$ selected events.  This is a very simple and qualitative model for the systematic error bars which follows the only purpose of avoiding unrealistic tiny error bars when only statistic uncertainties are taken into account.

We have compared the $m_{t\bar t}$-spectrum for the SM and the NP benchmark model when including the statistic and systematic uncertainties.  We have studied how these two spectrum differentiate as we perform single cuts only in $p_T(t\bar t)$ or $\beta$ and also in a combination of them, which in all cases increase the quark-fusion content in the selected sample.  

We first analyze the case of single cuts in $p_T(t\bar t)$ and $\beta$.  If we take the reference cuts in these variables to be $p_T(t\bar t)<20$ GeV and $\beta>0.85$ in each case respectively, one may qualitatively predict the results using Fig.~\ref{qqbarfraction}.  We see from the figure that in both cases we may expect a quark-fusion fraction of approximately $0.40$ in the selected sample.  This fraction comes from the value of the $\beta_{min}=0$ (red dot-dashed) line at $p_T(t\bar t)_{MAX}=20$ GeV and the limit of the $\beta_{min}=0.85$ (black solid) line for $p_T(t\bar t)_{MAX}\to \infty$.  On the other hand, we see that the case of only cutting in $p_T(t\bar t)$ keeps a $29$\% of the sample, whereas the case of only cutting in $\beta$ keeps only a $17$\% (not shown in the figure), henceforth we would expect that in a regime of statistic error dominance the cut in $p_T(t\bar t)$ would perform better since the statistic error increases less with the cut.  However, given the large luminosity collected insofar at the LHC, these percentages for the selected sample do not put in any risk a systematic error regime and, therefore, we may expect a similar behaviour of both cuts in enhancing the sensitivity of the $m_{t\bar t}$-spectrum. 
% There is, however, a possible extra source of sensitivity enhancement which has to do with which set of quark-fusion events induces each variable.  In fact, the selection using $\beta$ automatically selects quark-fusion events with larger invariant mass than those coming from the $p_T(t\bar t)$ selection.  Since NP effects are usually expected at larger invariant mass, a selection based on $\beta$ would be better in this case. 
 We have compared the SM and NP $m_{t\bar t}$-spectrum for the $\beta>0.85$ and $p_T(t\bar t)<20$ GeV selection methods through a $\chi^2$ test for $25$ GeV bins in $m_{t\bar t}$ in the $600$ to $800$ GeV region and found $p$-values of $p=0.63$ and $0.85$, respectively for each method.  We see that the $p$-values achieved with both cuts are statistically similar, as expected from previous arguments. We also see that none of both cuts is enough to differentiate NP from SM and, henceforth, they should be combined.   In addition, from the experimental point of view, it is also better to combine these variables rather than trying to reach a theoretical limit in any of them in order to increase the quark-fusion fraction.

\begin{figure}[!htb]
\begin{center}
\hspace*{-1.4cm}
\begin{minipage}[b]{0.42\linewidth}
\begin{center}
\includegraphics[width=1\textwidth]{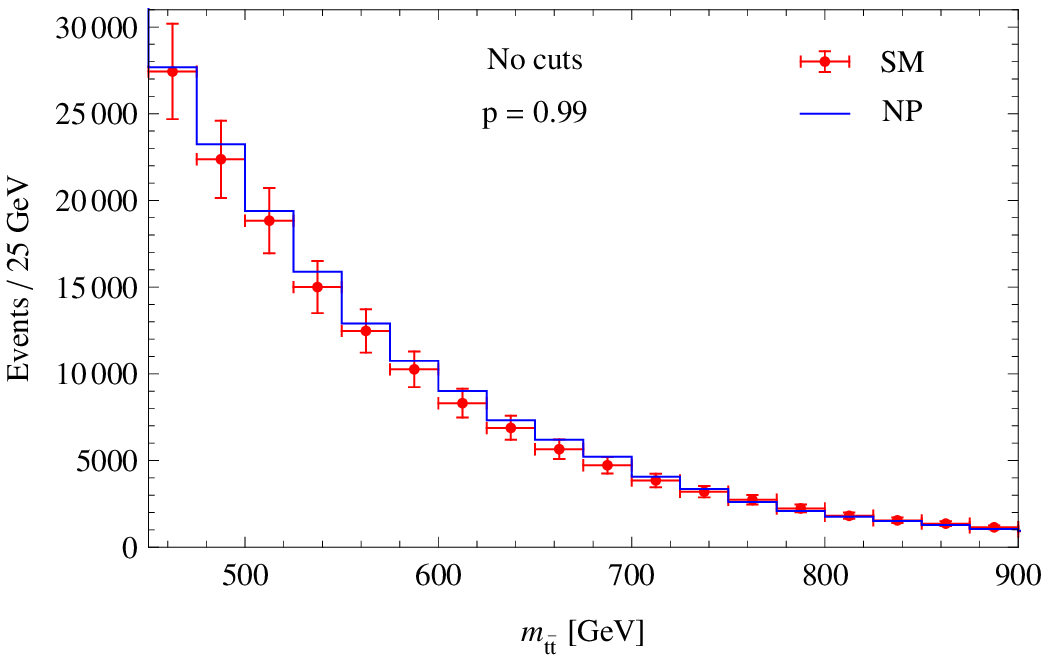}
\newline
{\tiny (a)}
\end{center}
\end{minipage}
\hspace{.3cm}
\begin{minipage}[b]{0.42\linewidth}
\begin{center}
\includegraphics[width=1\textwidth]{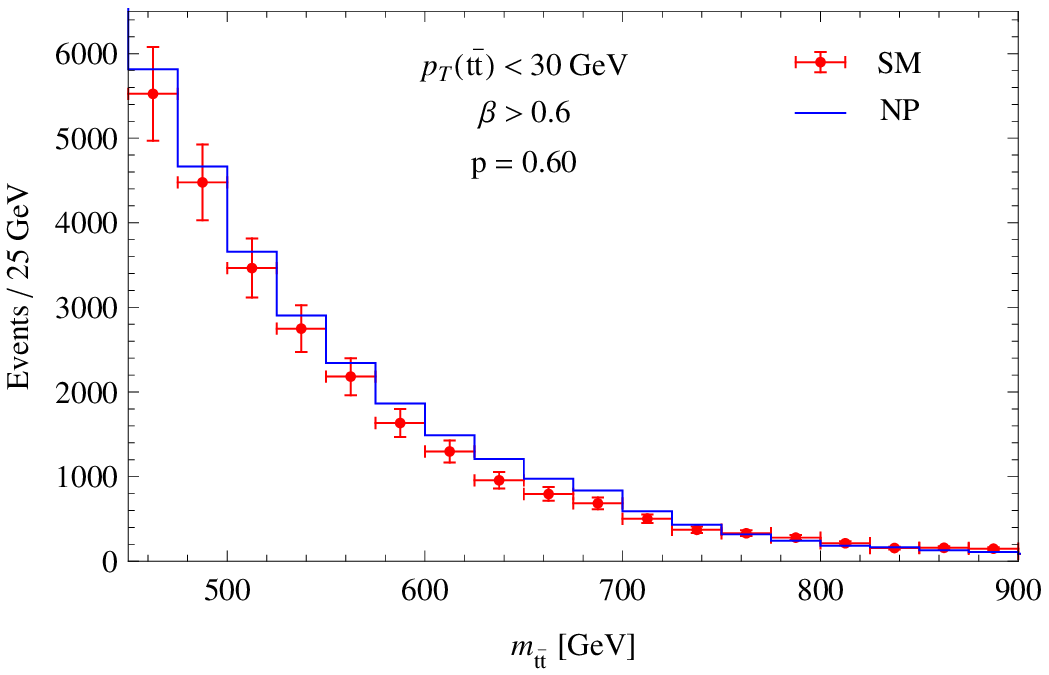}
\newline
{\tiny (b)}
\end{center}
\end{minipage}
\end{center}
\begin{center}
\hspace*{-1.4cm}
\begin{minipage}[b]{0.42\linewidth}
\begin{center}
\includegraphics[width=1\textwidth]{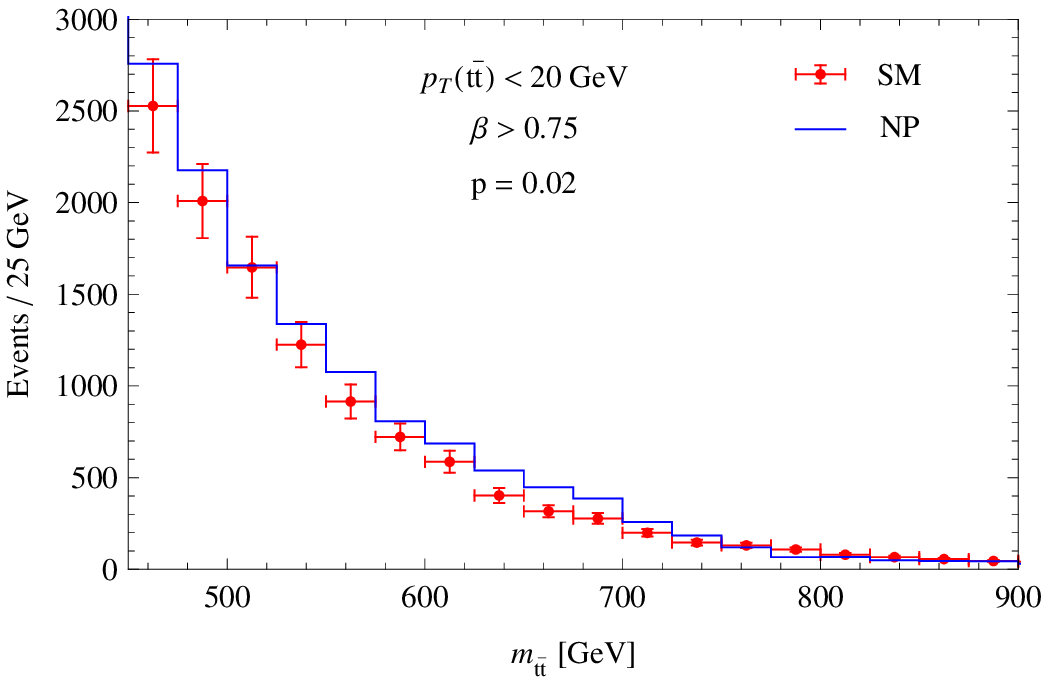}
\newline
{\tiny (c)}
\end{center}
\end{minipage}
\hspace{.3cm}
\begin{minipage}[b]{0.42\linewidth}
\begin{center}
\includegraphics[width=1\textwidth]{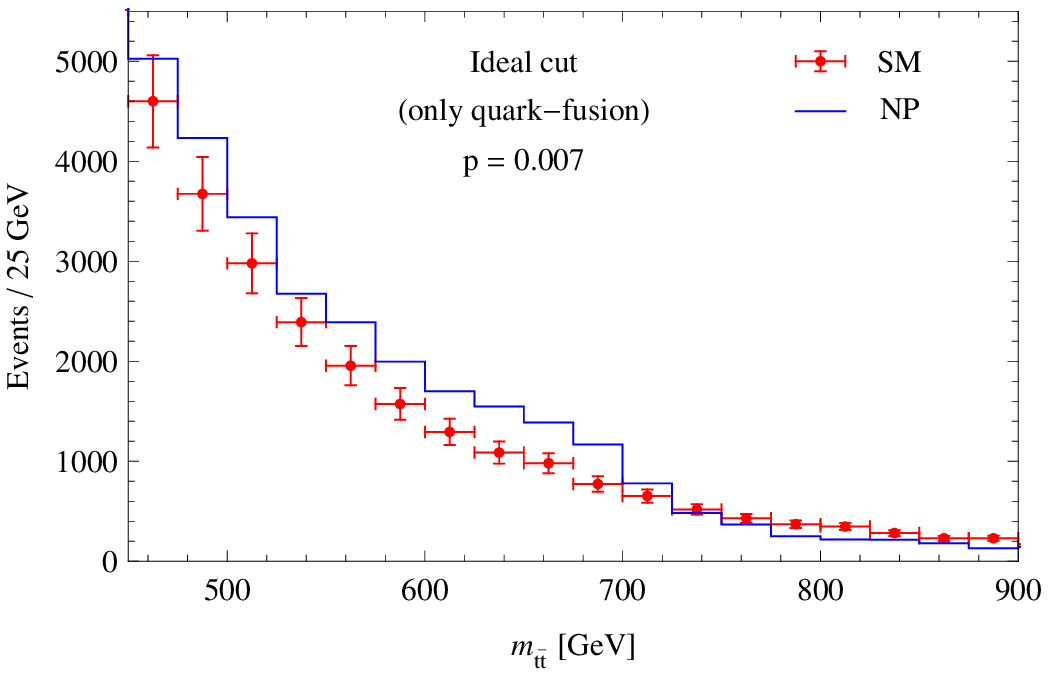}
\newline
{\tiny (d)}
\end{center}
\end{minipage}
\end{center}
\caption{Comparison of SM and NP $t\bar t$ invariant mass distributions for different selection cuts (a-c) and for the ideal cut in which only quark-fusion events are left (d).  The NP corresponds to a massive gluon partner with $m_{G'}=700$ GeV and width $\Gamma_{G'} = 90$ GeV.  The quoted $p$-values correspond to a $\chi^2$ test between 600 and 800 GeV.}
\label{comparison}
\end{figure}

We have performed combined cuts in $p_T(t\bar t)$ and $\beta$ to increase still more the quark-fusion content of the selected sample and, therefore, enhance the sensitivity to NP in the $m_{t\bar t}$ spectrum.  In Fig.~\ref{comparison} we have plotted both SM and NP spectrum with the total error bars on the SM expected spectrum for different combined cuts in $p_T(t\bar t)$ and $\beta$.  The last plot ($d$) shows the ideal cut in which only the quark-fusion events are left in the sample which, of course, is impossible to perform in the practice.    From Figs.~\ref{comparison}$a$-\ref{comparison}$c$ we see the improvement in the visibility of the resonant NP peak.  We have performed the same $\chi^2$ test as in the previous paragraph for the plots in Figs.~\ref{comparison}a-\ref{comparison}c and found that in this case the $p$-value improves from $p=0.99$ to $p=0.02$ thanks to the selection cuts. We have checked that if the systematic error bars are reduced below $20\%$  then the enhancement of the NP signal increases considerably.  
%If the error bars are above $20\%$ then the enhancement decreases.

From the results in Fig.~\ref{comparison} --and their quantification through the corresponding $p$-values-- we conclude that we could expect a considerable enhancement in the sensitivity to NP signals in the $m_{t\bar t}$-spectrum if the sample is correctly selected.  A final realistic evaluation in this enhancement could only be accomplished by the experimental groups, which have a full estimation of their systematic error bars.  Only a realistic estimation of the systematic error bars will allow experimentalists to optimize the weighting of the cuts in the $p_T(t\bar t)$ and $\beta$  variables.  Finally, an improvement in the low $p_T(t\bar t)$ spectrum of the Monte-Carlo generators by theoretical calculations may be required to further reduce the error bars.  

The enhancement in the sensitivity to NP signals in the $m_{t\bar t}$-spectrum found in this work using the $p_T(t\bar t)$ and $\beta$ variables is in consonance with the improvement in the induced $t\bar t$ asymmetries found in Ref.~\cite{yo} using the same variables.

The analysis presented in this work follows the purpose of predicting a possible improvement in the sensitivity to NP in the $m_{t\bar t}$-spectrum, not to quantify it.  The underlying idea of this study could also be used in the analysis of the LHC $b\bar b$-spectrum and related observables.  On the other hand, the enrichment in quark-fusion of the sample in dijet searches for NP is more involved, since in this case the final quarks could also be in the initial state and, therefore, the $t$-channel has a large contribution to the production cross-section \cite{lqv}.  The study in this work could be reversed for the case of NP coupling only to gluons, although the enhancement will not be so important.

\section*{Acknowledgments}
Thanks JoAnne Hewett and Stefan Hoeche for useful conversations. Thanks CONICET for the special funding and SLAC for its hospitality.

%%%%%%%%%%%%%%%%%%%%%%%%%%%%%%%%%%%%%%%%%%%%%%%%%%%%%%%%%%%%%%%%
{}

\end{document}